\documentclass{aa520}
\usepackage{graphicx}
\begin{document}
   \title{Direct detection of the companion of $\chi^1$ Orionis}

   \subtitle{}

   \author{B. K\"onig
          \inst{1}
          \and
          K. Fuhrmann\inst{1}
          \and
          R. Neuh\"auser\inst{1}
          \and 
          D. Charbonneau\inst{2} 
          \and
           R. Jayawardhana\inst{3}
          }

   \offprints{Brigitte K\"onig, \email{bkoenig@mpe.mpg.de}}

   \institute{Max-Planck-Institut f\"ur extraterrestrische Physik,
              Gie\ss enbachstra\ss e 1, D-85748 Garching, Germany
                  \and
              California Institute of Technology, 105-24 (Astronomy)
              1200 E. California Blvd., Pasadena CA 91125, U.S.A.
              	  \and
              Department of Astronomy, University of Michigan, Ann Arbor, MI
              48109, U.S.A. 
	                   }

   \date{28. August 2002; 19. September 2002}

   \abstract{
     We present an H-band image of the companion of $\chi^1$~Orionis
     taken with the Keck adaptive optic system and NIRC~2 camera equipped
     with a 300\,mas-diameter coronographic mask. The direct detection of
     this companion star enables us to calculate dynamical masses using only 
     Kepler's laws (M$_{\rm A} =1.01\pm0.13\,{\rm M}_{\odot}$, M$_{\rm B}
     =0.15\pm0.02\,{\rm M}_{\odot}$), and to study stellar evolutionary models
     at a wide spread of masses. The application of \cite{Baraffe1998}
     pre-main-sequence models implies an age of 70-130\,Myrs. This is in
     conflict to the age of the primary, a confirmed member of the Ursa Major
     Cluster with a canonical age of 300\,Myrs. As a consequence, either the
     models at low masses underestimate the age or the Ursa Major Cluster is
     considerably younger than assumed.
   }

   \maketitle
%

\section{Introduction}
$\chi^1$~Ori is a G0V-star and is known to be a single-lined spectroscopic and
astrometric binary. The orbital parameters were first derived 
by \cite{Lipp1978}. Since then \cite{Irvin1992} published precise
radial-velocity measurements of the orbit. \cite{Gatewood1994} published an 
astrometric parallax of the orbit of $\chi^1$~Ori. Recently, \cite{Han2002}
using their new astrometric data and the radial velocity data from
\cite{Marcy1998} presented a period of ${\rm P} = 5156.7 \pm 
2.5$~days and a mass ratio $q = {\rm M}_{\rm B}/{\rm M}_{\rm A} = 0.15 \pm
0.005$.\\ 
\begin{figure}[h]
{\includegraphics[angle=0, width=8.5cm]{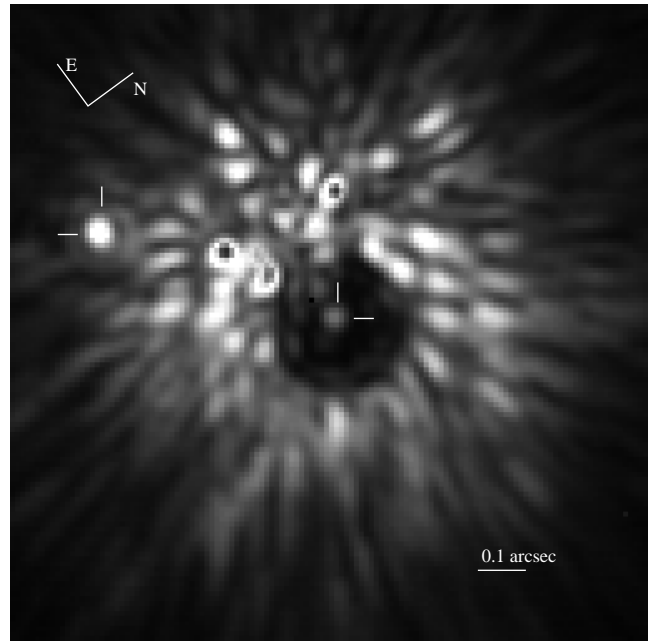}}
\caption{The H-band image of $\chi^1$~Ori behind the coronograph in the center
  and the companion to the left. Note the diffraction ring around the
  companion.}
\label{fig:image}
\vspace{-0.5cm}
\end{figure}
\cite{McCarthy1986} claimed to have detected the companion directly by speckle
imaging techniques, but this has not been confirmed yet. They derive 
M$_{\rm V}=6.1$\,mag, which would place the companion star to $\chi^1$~Ori
about 4\,mag above the main sequence (\cite{Henry1999}). \cite{Han2002} claim
that \cite{McCarthy1986} and subsequent attempts by speckle observations have
not been able to detect the companion directly due to instrument
limitations.\\ 
The G-type star $\chi^1$~Ori and its companion form a binary with a very small
mass ratio. A direct detection of the secondary would be significant as it
would allow the masses to be determined without astrophysical assumption. The
derived mass and observed luminosity allow the age to be inferred from
comparison to pre-main-sequence evolutionary tracks, which in turn enables a
calibration of other alternate estimators. 


\section{Data reduction and analysis}
We observed $\chi^1$~Ori on Feb. 28, 2002, using the Keck~2 telescope
equipped with the NIRC2 camera and the adaptive optic system
(\cite{Wizinowich2000}), an H-band filter and a 300\,mas diameter
coronograph. The coronograph is semi-transparent with a throughput slightly
below half a percent as determined by us (different from what is given in 
the manual, but confirmed by the Keck staff), so the position of the star
behind it can be measured precisely. The total integration time was
0.18\,s. The FWHM of the companion is 50\,mas.\\  
We performed the data reduction, using the reduction software MIDAS
(\cite{MIDAS}) provided by ESO. We divided the image by a normalized 
master sky-flat. We subtracted the background of $\chi^1$~Ori~B depending on
the distance from $\chi^1$~Ori~A: $\chi^1$~Ori~B is located in the PSF wing
of the component A which causes the main contribution to the background
emission. To obtain a background subtracted instrumental magnitude of B we
subtract an azimuthal averaged background (a one pixel wide annulus around A
excluding B) for each pixel used.\\
We determined the magnitude of the companion as well as that of two stars
used as photometric standards: the UKIRT faint standard FS~11, ${\rm H} =
11.276 \pm 0.003$\,mag (\cite{Hawarden2001}) and TWA-5B, ${\rm H} = 12.14 \pm
0.06$\,mag (\cite{Lowrance1999}). The standards were observed in the same
night, and analyzed with the same procedure. We used 121 different aperture
sizes starting with the brightest central pixel and calculating a background
subtracted peak-to-peak flux ratio and then consecutively adding the next
brightest pixel until we end up with a $121 = 11 \times 11$~pixel aperture
box. For aperture sizes from 1 to 50 pixels, the resulting instrumental
magnitude did not change significantly, so we use this value. By comparing the
background subtracted instrumental magnitude of the companion to the
background subtracted instrumental magnitudes obtained for TWA-5~B
and FS~11, we measure the apparent H-band magnitude for the $\chi^1$~Ori
companion of $7.70 \pm 0.15$\,mag, taking into account also the slightly
different FWHM and Strehl ratios. With the Hipparcos parallax for
$\chi^1$~Ori of $115.43\pm1.08$\,mas, we obtain ${\rm M}_{\rm H} = 8.01 \pm
0.15$\,mag for the B-component.  

\section{Dynamical masses of $\chi^1$~Orionis A \& B}
Using the orbital elements for $\chi^1~{\rm Ori}$~published by 
\cite{Han2002} ($i = 95.937 \pm 0.790$\,deg, $T_0({\rm JD}) = 2,451,468.2 \pm
3.083$, $P = 5156.291 \pm 1.508$\,days, $e = 0.452 \pm 0.002$), we can
calculate the absolute masses of $\chi^1~{\rm Ori}$~A and B directly using
Kepler's laws. The measured apparent separation between the 
components A and B is $\rho = 0.4976 \pm 0.0036$\,arcsec using a
pixel-scale of $0.009942 \pm 0.000500$\,arcsec/pixel as determined by the NIRC
2 team (Campbell, priv. comm.). The physical separation then is $r=4.33 \pm
0.08$\,AU. This results in a mass for the primary of M$_{\rm A} = 1.01 \pm
0.13\,{\rm M}_{\odot}$ and for the secondary of M$_{\rm B} = 0.15
\pm 0.02\,{\rm M}_{\odot}$. Using the parallax determined by
\cite{Han2002} of $\chi^1$~Ori of $115.69 \pm 0.74$\,mas, the masses would be
M$_{\rm A} = 1.02 \pm 0.08$\,M$_\odot$ and  M$_{\rm B} = 0.15 \pm
0.01$\,M$_\odot$. The error-bars are fairly large but further direct 
measurements will improve the orbital solution, in particular the separation
and the position of the orbit in the sky. The position angle between
$\chi^1~{\rm Ori}$~A and B on the observing date (MJD = 52334.33952) is
$(123.22 \pm 0.12)^{\circ}$. The observed position and position angle is only
13\,mas and $2.8^\circ$~away from the predicted values by \cite{Han2002}.

\section{Spectral synthesis analysis of $\chi^1$~Orionis}
\begin{figure}[h]
{\includegraphics[angle=90, width=8.5cm]{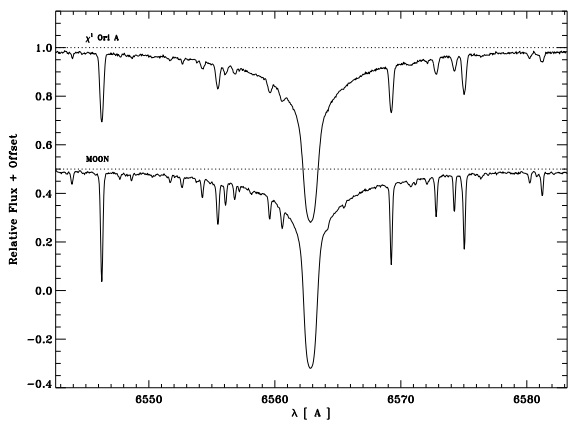}}
\caption{A spectrum of $\chi^1$~Ori~A compared to the moon (= reflected sun
  light) in the range of H$_{\alpha}$ at 6563\,\AA. Note the considerable
  rotational velocity of $\chi^1$~Ori~A, $v \sin i = 8.7$\,km/s, and the
  slightly filled-in line core of H$_{\alpha}$.}
\label{fig:Ha}
\end{figure}

\begin{figure}[h]
{\includegraphics[angle=90, width=8.5cm]{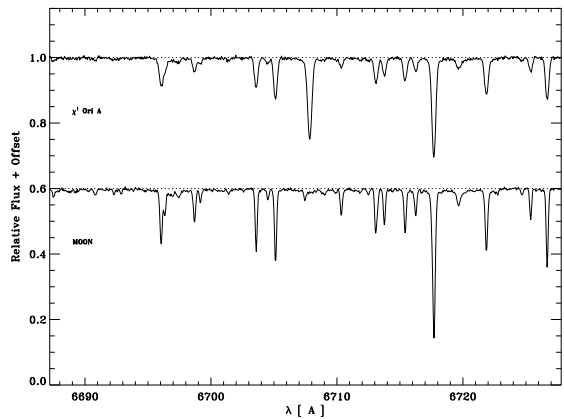}}
\caption{Same as Fig.~2, but for the range of
  lithium at 6707.8\,\AA~and calcium at 6717.8\,\AA.}
\label{fig:li}
\end{figure}

The basic stellar parameters of $\chi^1$~Ori~A are derived from a
model atmosphere analysis of high resolution, high S/N \'echelle
spectra (Fig.~\ref{fig:Ha} and \ref{fig:li}) obtained in January 2000 at the
Calar Alto Observatory, Spain, 2.2\,m telescope with FOCES
(\cite{Pfeiffer1998}). The fairly high projected rotational velocity $v \sin i
= 8.7\pm0.8$~km~s$^{-1}$, the strong lithium feature at $\lambda$6707
(Fig.~\ref{fig:li}), the "dipper-star-like" kinematics
($U/V/W=24/7/0$~km~s$^{-1}$), and the filled-in line cores of H$_\alpha$ 
(Fig.~\ref{fig:Ha}) and the Ca~II infrared triplet all consistently confirm
that $\chi^1$~Ori must belong to the Ursa Major Cluster. As in
\cite{Fuhrmann1997} we deduce the effective temperature of the primary,
$T_{\rm eff}=5920\pm70~$K, from the Balmer line wings and the surface gravity,
$\log g = 4.39\pm0.10$, from the iron ionization equilibrium and the wings of
the Mg~Ib lines. We find the metallicity to be slightly below the solar value
([Fe/H]$=-0.07\pm0.07$), again very typical for the mean abundance of Ursa
Major Cluster stars of $\langle [{\rm Fe/H}] \rangle = -0.09$
(\cite{Boesgaard1990}).  
With a bolometric magnitude M$_{\rm bol}=4.60\pm0.05$, and $T_{\rm eff}$ and
[Fe/H] as derived above, we find the mass to be
M$=1.04~{\rm M}_{\odot}$ (implied from the tracks given in Fuhrmann et
al. 1997), i.e. slightly above solar and with an uncertainty of 
about $0.05~{\rm M}_{\odot}$. The secondary -- being more than five magnitudes
(extrapolating the measured H-band magnitude) fainter in the visible -- does
not have an impact on our spectra. 

\section{Results and Discussion}
\begin{figure}[h]
{\includegraphics[angle=270, width=8.5cm]{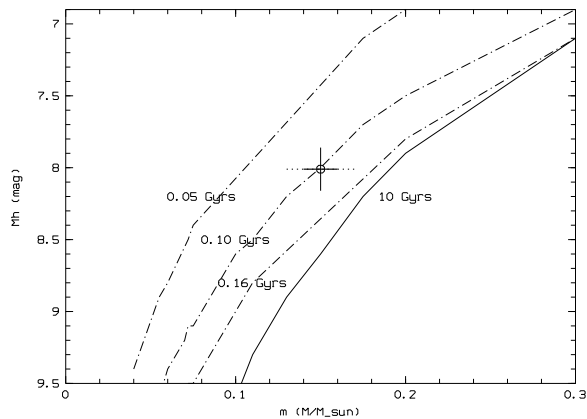}}
\caption{Baraffe et al. (1998) isochrones for solar metallicity in a
  mass-luminosity plot compared to the position of $\chi^1$~Ori~B. The
  error-bars for the mass are derived by the spectroscopy (solid)
  and for the dynamical mass (dots). The age for $\chi^1$~Ori~B ranges from
  70-130\,Myrs using the dynamical mass.}
\label{fig:m_mh}
\end{figure} 
\begin{figure}[ht]
\vspace{-0.5cm}
{\includegraphics[angle=90, width=9.5cm]{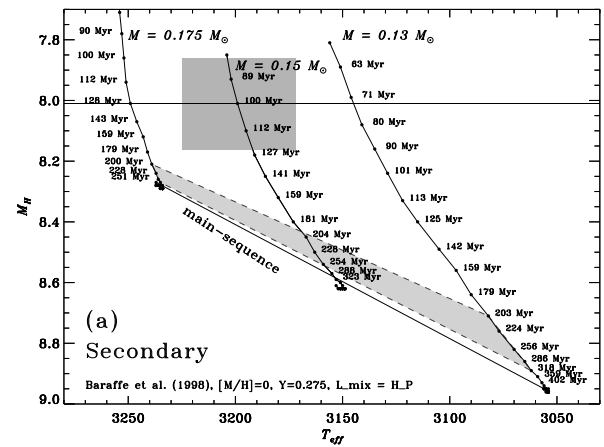}}
\vspace{-0.6cm}
{\includegraphics[angle=90, width=9.5cm]{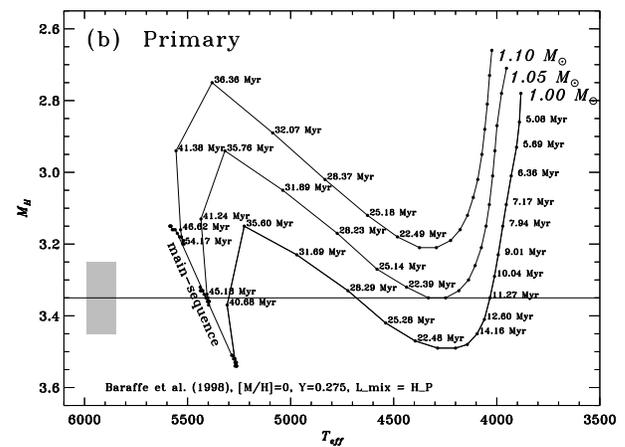}}
\vspace{-0.6cm}
{\includegraphics[angle=90, width=9.5cm]{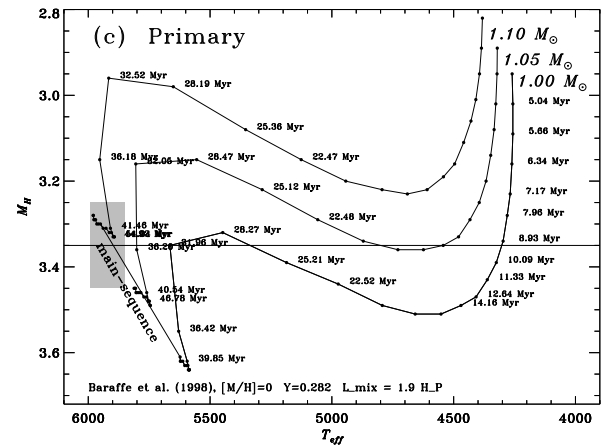}}
\caption{Baraffe et al. (1998) tracks for solar metallicity. The horizontal
   line in the first plot gives M$_{\rm H}$ for the companion star with the
   top shaded area indication the $1 \sigma$~error for M$_{\rm H}$ and the
   temperature range. In panel (a), the bottom shaded area is the age range
   determined for the Ursa Major cluster using different methods. With a mass
   of $0.15\,{\rm M}_{\odot}$ the companion appears younger compared to the age
   range of the Ursa Major cluster. In the other two panels the same tracks
   plotted are for the primary, indicating the position of the primary by the
   shaded area. In panel (a) and (b) the model parameters are [M/H]=0, Y=0.275
   and $L_{\rm mix} = H_{\rm P}$. For (c) the parameters have been adjusted to
   fit the sun to [M/H]=0, Y=0.282 and $L_{\rm mix} = 1.9 H_{\rm P}$.} 
\label{fig:teff_mh}
\end{figure}
The mass of the companion to $\chi^1$~Ori has been determined precisely to
$(0.15\pm0.005)\,{\rm M}_{\chi^1~{\rm Ori}}$ (Han \& Gatewood 2002). The
main uncertainty is ${\rm M}_{\chi^1~{\rm Ori}}$. This leads to a
spectroscopic ($0.15 \pm 0.01\,{\rm M}_{\odot}$) and dynamic mass ($0.15 \pm
0.02\,{\rm M}_{\odot}$), which are both in good agreement.\\
The position of the  $\chi^1$~Ori~B in the mass-luminosity plot
(Fig.~\ref{fig:m_mh} and \ref{fig:teff_mh}a) compared to the isochrones
provided by \cite{Baraffe1998} indicates that the star lies about
$0.50\pm0.10$\,mag above the main sequence.\\ 
Figure~\ref{fig:teff_mh}b and c show H-R diagrams for the primary star
including the tracks of \cite{Baraffe1998}. Figure~\ref{fig:teff_mh}b 
shows models for $[M/H] = 0$, $Y=0.275$, and the mixing length of $L_{\rm mix}
= H_{\rm P}$. \cite{Baraffe1998} acknowledge that these models do not
reproduce the sun at present age. Those tracks and isochrones also do not
reproduce $\chi^1$~Ori~A.\\ 
Figure~\ref{fig:teff_mh}c shows the same as Fig.~\ref{fig:teff_mh}b
except that the parameters $[M/H] = 0$, $Y=0.282$, and the mixing length of
$L_{\rm mix} = 1.9\,H_{\rm P}$ were adjusted to fit the sun. With these
parameters the present sun could be reproduced and for $\chi^1$~Ori~A they
also seem to work. The M$_{\rm H}$~predicted by \cite{Baraffe1998} is a bit
lower than the measured M$_{\rm H}$~value for $\chi^1$~Ori~A. This could be
because $\chi^1$~Ori~A is slightly iron underabundant (${\rm [Fe/H]} =
-0.07\pm0.07$) and the tracks were calculated for solar abundance. No tracks
for masses of 0.15 - 0.175\,M$_{\odot}$ are available for the model with the
parameter set to fit the sun.\\  
The age prediction by the pre-main-sequence models can be directly compared to
other age determinations for the Ursa Major Cluster.
While the canonical value for the age of the Ursa Major Cluster
is 300\,Myrs (cf. e.g. \cite{Soderblom1993}, and references therein) derived
by comparing the members of the Ursa Major Cluster nucleus stars in a
color-magnitude diagram to theoretical isochrones computed by
\cite{VandenBerg1985}, more recent observations of Sirius' white dwarf
companion led \cite{Holberg1998} to suggest an age of 160\,Myrs with
reference to the cooling tracks of \cite{Wood1992}. Since Sirius~B is also
well-known as a fairly massive degenerate white dwarf with a mass of ${\rm
  M}=1.034\pm0.026~{\rm M}_{\odot}$ (Holberg et al. 1998), the initial-final
mass relation suggests a progenitor of about 6-7~${\rm M}_{\odot}$ which means
that we can expect another $\sim$60-70~Myr for the pre-white-dwarf
evolution. Hence, an age only somewhat above 200\,Myrs may be more in line
with this nearby open cluster. More recent white dwarf cooling models of
\cite{Salaris2000} (models with a pure hydrogen atmosphere) suggest the age of
the white dwarf of 111\,Myrs derived 
from the V-magnitude and the temperature published by
\cite{Holberg1998}. Assuming the lifetime of the progenitor of the white dwarf
of 46\,Myrs this leads to an age of the UMa cluster of 157\,Myrs. \\
The comparison of the age using \cite{Baraffe1998} (70-130\,Myrs) to the ages
of the Ursa Major Cluster (200-300\,Myrs) indicate that either: (i) the  Ursa
Major Cluster has a larger than expected age spread, (ii) there are problems
with the models at a solar and/or at $\sim 0.15\,{\rm M}_{\odot}$~mass, (iii)
the canonical age for the Ursa Major Cluster is too high (300\,Myrs), or (iv)
$\chi^1$~Ori is not a member of the Cluster. 
Considering possibility (i), we note that the age
spread of 70-300\,Myrs seems 
too large for a Cluster.  As for the option (iv),  $\chi^1$~Ori is
a classical member of the Ursa Major Cluster, located near the cluster
center. The spectrum of $\chi^1$~Ori~A would support an
age of 200\,Myrs regarding the activity indicators, as would the cooling
tracks for the Sirius B white dwarf.

\begin{acknowledgements}
This research has made use of the SIMBAD database, operated at CDS,
Strasbourg, France. B.K. wants to thank F. Dufey for help with the algebra.  
R.N. wishes to acknowledge financial support from the Bundesministerium f\"ur
Bildung und Forschung through the Deutsches Zentrum f\"ur Luft- und Raumfahrt
e.V. (DLR) under grant number 50 OR 0003. R.J. wishes to acknowledge support
from NASA grant NAG5-11905. Some of the Data presented herein
were obtained at the W.M. Keck Observatory, which is operated as a scientific
partnership among the California Institute of Technology, the University of
California and the National Aeronautics and Space Administration. The
Observatory was made possible by the generous financial support of the
W.M. Keck Foundation. The authors wish to recognize and acknowledge the very
significant cultural role and religious significance that the summit of Mauna
Kea has always had within the indigenous Hawaiian community.  We are most
fortunate to have the opportunity to conduct observations from this mountain. 
The Authors would like to thank Randy Campbell and David LeMignant for help
during the observing nights. We thank the referee, G. Gatewood for the helpful
comments. 
\end{acknowledgements}

\end{document}